\documentclass[%
 reprint,
superscriptaddress,
nofootinbib,
nobibnotes,
 amsmath,amssymb,
 aps,
]{revtex4-1}

\usepackage{graphicx}% Include figure files
\usepackage{dcolumn}% Align table columns on decimal point
\usepackage{bm}% bold math
\usepackage{multirow}
\usepackage{color}
\usepackage{amsmath}
\usepackage{mathtools}
\def\l{\left(}
\def\r{\right)}

\begin{document}

\preprint{APS/123-QED}

%\title{Manuscript Title:\\with Forced Linebreak}% Force line breaks with \\
\title{Search for the rare decays ${\pi}^+{\to}\mu^+{\nu}_\mu{\nu}\bar{\nu}$ and ${\pi}^+{\to}e^+{\nu}_e{\nu}\bar{\nu}$ } % Force line breaks with \\
%\thanks{A footnote to the article title}%

\author{A.~Aguilar-Arevalo}
\affiliation{Instituto de Ciencias Nucleares, Universidad Nacional Aut\'onoma de M\'exico, CDMX 04510, M\'exico}

\author{ M.~Aoki}
\affiliation{Physics Department, Osaka University, Toyonaka, Osaka, 560-0043, Japan}

\author{M.~Blecher}
\affiliation{Virginia Tech., Blacksburg, Virginia 24061, USA}

\author{D.I.~Britton}
\affiliation{SUPA - School of Physics and Astronomy, University of Glasgow, Glasgow, G12-8QQ, United Kingdom}

\author{D.~vom~Bruch}
\thanks{Present address: LPNHE, Sorbonne Universit\'e, Universit\'e Paris Diderot, CNRS/IN2P3, Paris, France.}
\affiliation{Department of Physics and Astronomy, University of British Columbia, Vancouver, British Columbia V6T 1Z1, Canada}

\author{D.A.~Bryman}
\affiliation{Department of Physics and Astronomy, University of British Columbia, Vancouver, British Columbia V6T 1Z1, Canada}
\affiliation{TRIUMF, 4004 Wesbrook Mall, Vancouver, British Columbia V6T 2A3, Canada}

\author{S.~Chen}
\affiliation{Department of Engineering Physics, Tsinghua University, Beijing, 100084, China}

\author{J.~Comfort}
\affiliation{Physics Department, Arizona State University, Tempe, AZ 85287, USA}

\author{S.~Cuen-Rochin}
%\affiliation{Department of Physics and Astronomy, University of British Columbia, Vancouver, B.C., V6T 1Z1, Canada}
\affiliation{TRIUMF, 4004 Wesbrook Mall, Vancouver, British Columbia V6T 2A3, Canada}
\affiliation{Universidad Aut\'onoma de Sinaloa, Culiac\'an, M\'exico}

\author{L.~Doria}
\affiliation{TRIUMF, 4004 Wesbrook Mall, Vancouver, British Columbia V6T 2A3, Canada}
\affiliation{PRISMA$^+$ Cluster of Excellence and Institut f\"ur Kernphysik, Johannes Gutenberg-Universit\"at Mainz, Johann-Joachim-Becher-Weg 45, D 55128 Mainz, Germany}

\author{P.~Gumplinger}
\affiliation{TRIUMF, 4004 Wesbrook Mall, Vancouver, British Columbia V6T 2A3, Canada}

\author{A.~Hussein}
\affiliation{TRIUMF, 4004 Wesbrook Mall, Vancouver, British Columbia V6T 2A3, Canada}
\affiliation{University of Northern British Columbia, Prince George, British Columbia V2N 4Z9, Canada}

\author{Y.~Igarashi}
\affiliation{KEK, 1-1 Oho, Tsukuba-shi, Ibaraki, 300-3256, Japan}

\author{S.~Ito}
\thanks{Corresponding author (s-ito@okayama-u.ac.jp).\\Present address: Faculty of Science, Okayama University, Okayama, 700-8530, Japan.}
\affiliation{Physics Department, Osaka University, Toyonaka, Osaka, 560-0043, Japan}

\author{S.~Kettell}
\affiliation{Brookhaven National Laboratory, Upton, NY, 11973-5000, USA}

\author{L.~Kurchaninov}
\affiliation{TRIUMF, 4004 Wesbrook Mall, Vancouver, British Columbia V6T 2A3, Canada}

\author{L.S.~Littenberg}
\affiliation{Brookhaven National Laboratory, Upton, NY, 11973-5000, USA}

\author{C.~Malbrunot}
\thanks{Present address: Experimental Physics Department, CERN, Gen\`eve 23, CH-1211, Switzerland.}
\affiliation{Department of Physics and Astronomy, University of British Columbia, Vancouver, British Columbia V6T 1Z1, Canada}

\author{R.E.~Mischke}
\affiliation{TRIUMF, 4004 Wesbrook Mall, Vancouver, British Columbia V6T 2A3, Canada}

\author{T.~Numao}
\affiliation{TRIUMF, 4004 Wesbrook Mall, Vancouver, British Columbia V6T 2A3, Canada}

\author{D.~Protopopescu}
\affiliation{SUPA - School of Physics and Astronomy, University of Glasgow, Glasgow, United Kingdom}

\author{A.~Sher}
\affiliation{TRIUMF, 4004 Wesbrook Mall, Vancouver, British Columbia V6T 2A3, Canada}

\author{T.~Sullivan}
\thanks{Present address: Department of Physics, University of Victoria, Victoria BC V8P 5C2, Canada.}
\affiliation{Department of Physics and Astronomy, University of British Columbia, Vancouver, British Columbia V6T 1Z1, Canada}

\author{D.~Vavilov}
%\affiliation{Department of Physics and Astronomy, University of British Columbia, Vancouver, British Columbia V6T 1Z1, Canada}
\affiliation{TRIUMF, 4004 Wesbrook Mall, Vancouver, British Columbia V6T 2A3, Canada}

\author{D. Gorbunov}
\affiliation{Theoretical Physics Division, Institute for Nuclear Research of the Russian Academy of Sciences, 60th October anniversary prospect, 7a, Moscow, 117312 Russia}
\affiliation{Moscow Institute of Physics and Technology, Institutsky lane 9, Dolgoprudny, Moscow region, 141700, Russia}

\author{D. Kalashnikov}
%\affiliation{Theoretical Physics Division, Institute for Nuclear Research of the Russian Academy of Sciences, 60th October anniversary prospect, 7a, Moscow, 117312 Russia}
\affiliation{Moscow Institute of Physics and Technology, Institutsky lane 9, Dolgoprudny, Moscow region, 141700, Russia}

%\altaffiliation[Also at ]{Physics Department, XYZ University.}%Lines break automatically or can be forced with \\
%\author{Second Author}%
% \email{Second.Author@institution.edu}
%\affiliation{%
% Authors' institution and/or address\\
% This line break forced with \textbackslash\textbackslash
%}%

\collaboration{PIENU Collaboration}%\noaffiliation

%\author{Charlie Author}
% \homepage{http://www.Second.institution.edu/~Charlie.Author}
%\affiliation{
% Second institution and/or address\\
% This line break forced% with \\
%}%
%\affiliation{
% Third institution, the second for Charlie Author
%}%
%\author{Delta Author}
%\affiliation{%
% Authors' institution and/or address\\
% This line break forced with \textbackslash\textbackslash
%}%

%\collaboration{CLEO Collaboration}%\noaffiliation

\date{\today}% It is always \today, today,
             %  but any date may be explicitly specified

\begin{abstract}

The rare pion decays ${\pi}^+{\to}\mu^+{\nu}_\mu{\nu}\bar{\nu}$ and ${\pi}^+{\to}e^+{\nu}_e{\nu}\bar{\nu}$   are allowed in the Standard Model but highly suppressed.  
These decays were searched for using data from the PIENU experiment. 
A first result for ${\Gamma}({\pi}^+{\to}{\mu}^+{\nu}_{\mu}{\nu}\bar{\nu})/{\Gamma}({\pi}^+{\to}{\mu}^+{\nu}_{\mu})<8.6{\times}10^{-6}$, and an improved measurement   ${\Gamma}({\pi}^+{\to}e^+{\nu}_e{\nu}\bar{\nu})/{\Gamma}({\pi}^+{\to}{\mu}^+{\nu}_{\mu})<1.6{\times}10^{-7} $ were obtained.

\end{abstract}

%\keywords{Suggested keywords}%Use showkeys class option if keyword
                              %display desired
\maketitle

%\tableofcontents

\section{\label{sec:Introduction}Introduction}

The standard model (SM) allows for second order leptonic four-body pion and kaon decays $\pi^+/K^+{\to}{\mu}^+{\nu_{\mu}}{\nu}\bar{\nu}$ and $\pi^+/K^+{\to}e^+{\nu_e}{\nu}\bar{\nu}$ where ${\nu}\bar{\nu}$ includes all three neutrino families ${\nu_e}\bar{\nu_e}$, ${\nu_{\mu}}\bar{\nu_{\mu}}$, and ${\nu_{\tau}}\bar{\nu_{\tau}}$. 
The latest calculation of these processes was performed by Gorbunov and Mitrofanov \cite{3nu} with SM predictions for the branching ratios of kaon decays of order of $10^{-16}$. 
Due to the high level of suppression, experimental investigation of these processes could reveal small non-SM effects such as neutrino-neutrino ($I_{{\nu}\bar{\nu}}$) interactions \cite{nunu} and six-fermion (6f) interactions \cite{6-fermion, 6-fermion2}, which might compete with the SM processes at first order.

The rare kaon decay $K^+{\to}{\mu}^+{\nu_{\mu}}{\nu}\bar{\nu}$ was first searched for by Pang {\it et al.} \cite{Pang} resulting in a 90\% confidence level  (C.L.)  upper limit\footnote{All subsequent limits will be presented at the 90 \% C.L..} on the branching ratio ${\Gamma}(K^+{\to}{\mu}^+{\nu_{\mu}}{\nu}\bar{\nu})/{\Gamma}(K^+{\to}{\rm all})<6{\times}10^{-6}$. 
The most recent experimental study for $K^+{\to}{\mu}^+{\nu_{\mu}}{\nu}\bar{\nu}$ decay was performed by Artamonov {\it et al}. \cite{E949} giving the upper limits on  the branching ratio $B^{K\mu3\nu}_{\rm SM}={\Gamma}(K^+{\to}{\mu}^+{\nu_{\mu}}{\nu}\bar{\nu})/{\Gamma}(K^+{\to}{\rm all})<2.4{\times}10^{-6}$ for the SM,  $B^{K\mu3\nu}_{I_{{\nu}\bar{\nu}}}<2.4{\times}10^{-6}$ for the neutrino-neutrino interaction model, and $B^{K\mu3\nu}_{\rm 6f}<2.7{\times}10^{-6}$ for the six-fermion interaction model. 
The decay $K^+{\to}e^+{\nu_e}{\nu}\bar{\nu}$ assuming the neutrino-neutrino interaction model was searched for by Heintze {\it et al.} \cite{Ke3nu} resulting in the upper limit on the branching ratio $B^{Ke3{\nu}}_{I_{{\nu}\bar{\nu}}}={\Gamma}(K^+{\to}e^+{\nu_e}{\nu}\bar{\nu})/{\Gamma}(K^+{\to}{\rm all})<6{\times}10^{-5}$. 
The rare pion decay ${\pi}^+{\rightarrow}e^+{\nu}_e{\nu}\bar{\nu}$ was searched for by Picciotto et al. \cite{Picciotto} using the positron energy spectrum from ${\pi}^+{\to}e^+{\nu}$ decay. 
The upper limit on the branching ratio assuming the SM was found to be $B^{\pi e3\nu}_{\rm SM}={\Gamma}({\pi}^+{\to}e^+{\nu_e}{\nu}\bar{\nu})/{\Gamma}({\pi}^+{\to}{\mu}^+{\nu_{\mu}})<5{\times}10^{-6}$. 

In the present work, the rare pion decays ${\pi}^+{\to}{\mu}^+{\nu_{\mu}}{\nu}\bar{\nu}$ and ${\pi}^+{\to}e^+{\nu_e}{\nu}\bar{\nu}$ were sought using the full data set of the PIENU experiment \cite{PIENU} performed from 2009 to 2012 corresponding to an order of magnitude larger statistics than the previous TRIUMF experiment \cite{Picciotto}. 
The analyses are based on the searches for heavy neutrinos ${\nu}_H$ in ${\pi}^+{\to}{\mu}^+{\nu_H}$ decay \cite{PIENU2} and ${\pi}^+{\to}e^+{\nu_H}$ decay \cite{PIENU3}. We also present new theoretical estimates  for the SM branching ratios for  ${\pi}^+{\to}{\mu}^+{\nu_{\mu}}{\nu}\bar{\nu}$ and ${\pi}^+{\to}e^+{\nu_e}{\nu}\bar{\nu}$ decays.

\section{Theory}

\subsection{The Standard Model weak interaction}
\begin{figure}
\includegraphics[scale=0.45]{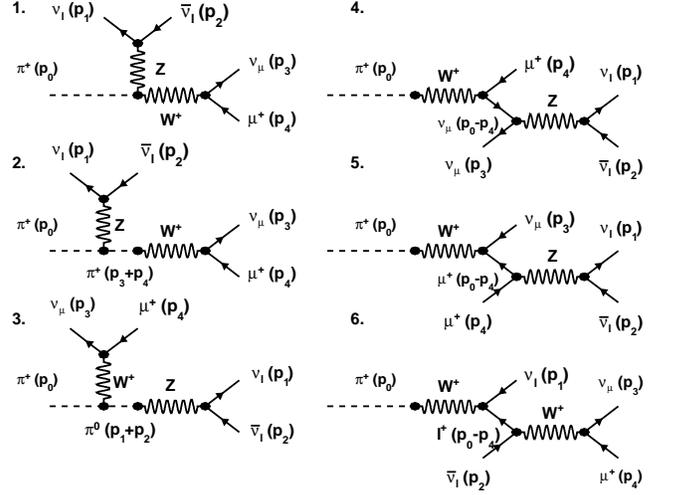}
\caption{\label{fig:Feynman} Feynman diagrams of ${\pi}^+{\to}{\mu}^+{\nu}_{\mu}{\nu}_l\bar{\nu}_l$ decay processes in the SM framework where $p_i$ represents the momentum of the $i^{th}$ particle and 
$l=e,\mu,\tau$. Similar diagrams contribute to ${\pi}^+{\to}e^+{\nu}_e{\nu}_l\bar{\nu}_l$ decay.}
\end{figure}

The SM second order decay rates for  ${\pi}^+{\to}{\mu}^+{\nu_{\mu}}{\nu}\bar{\nu}$ and ${\pi}^+{\rightarrow}e^+{\nu_e}{\nu}\bar{\nu}$ were estimated in the framework of Chiral Perturbation theory (ChPT) using the procedures of Gorbunov and Mitrofanov \cite{3nu} which were used for the analysis of the equivalent $K$ decays \cite{E949}. Fig. \ref{fig:Feynman} shows the relevant Feynman diagrams for  ${\pi}^+{\to}{\mu}^+{\nu_{\mu}}{\nu}\bar{\nu}$ decay.
Each of the three neutrino generations, calculated separately, contributes to the ${\nu}\bar{\nu}$ pairs in the final state and to the combined charged lepton spectra. 
The branching ratios were calculated to leading order $O(p^2)$ in the momentum expansion as in Ref. \cite{3nu} except for using the pion mass, the pion decay constant $f_{\pi}$, the quark coupling $V_{ud}$, and the appropriate phase space. 
The relevant interaction terms originated from the ChPT $O(p^2)$ Lagrangian and the leptonic weak current 
part of the SM Lagrangian can be represented as:
\begin{eqnarray}\label{eq:sm}
\mathcal{L} &=& \frac{i f_{\pi} g^2 \sin^2 \theta_W }{2 \cos \theta_W} V_{ud} Z^{\mu} W^-_{\mu} \pi^+ \!-\! \frac{f_{\pi} g}{2} V_{ud}W^-_{\mu} \partial^{\mu} \pi^+ \nonumber \\ 
&& \!+\! ig \frac{2 \sin^2 \theta_W - 1}{2 \cos \theta_W} Z^{\mu} \!\l\partial_{\mu} \pi^+ \pi^- \!-\!  \pi^+\partial_{\mu} \pi^-\r  \nonumber \\ 
&&+ \frac{i g W_{\mu}^- V_{ud}}{2} \left[\partial_{\mu} \pi^+ \pi^0
- \pi^+ \partial_{\mu} \pi^0 \right] - \frac{g f_{\pi}Z_{\mu}}{2  \cos \theta_W} \times\partial_{\mu} 
\pi^0 \nonumber \\
&&-\frac{g}{2 \sqrt 2} \l W_{\mu}^+ \bar \nu_l \gamma^{\mu} (1 - \gamma_5) l + h.c. \r \nonumber \\ 
&&-\frac{g Z_{\mu}}{4 \cos \theta_W} \bar \nu_l \gamma^{\mu} (1 - \gamma_5) \nu_l
\end{eqnarray}
where $g$ is the coupling constant defined by the Fermi coupling constant $G_F=g^2/4\sqrt{2}M^2_W$ and  $M_W$ is the mass of W boson. 

Figure \ref{fig:EnergySpectra} shows the muon kinetic energy ($T_{\mu}$) spectrum for ${\pi}^+{\to}{\mu}^+{\nu_{\mu}}{\nu}\bar{\nu}$ decays and the positron total energy ($E_e$) spectrum for ${\pi}^+{\rightarrow}e^+{\nu_e}{\nu}\bar{\nu}$ decays. 
The results for the branching ratios for ${\pi}^+{\to}{\mu}^+{\nu_{\mu}}{\nu}\bar{\nu}$ and ${\pi}^+{\rightarrow}e^+{\nu_e}{\nu}\bar{\nu}$ decays in the SM framework were found to be $B^{\pi\mu3\nu}_{\rm SM}=4.0\times10^{-20}$ and $B^{\pi e 3\nu}_{\rm SM}=1.7\times10^{-18}$, respectively. 
Details of the SM theory calculations are given in the Appendix. 

\subsection{Non-Standard Model interactions}

Using the model suggested by Bardin, Bilenky, and Pontecorvo incorporating non-SM interactions between neutrinos \cite{nunu}, the differential decay rate for $K^+{\to}{\mu}^+{\nu_{\mu}}{\nu}\bar{\nu}$ decay was calculated in Refs. \cite{nunu} and \cite{Pang}. 
The energy spectrum for the neutrino-neutrino interaction model for the pion decay ${\pi}^+{\to}{\mu}^+{\nu_{\mu}}{\nu}\bar{\nu}$ (${\pi}^+{\rightarrow}e^+{\nu_e}{\nu}\bar{\nu}$) was obtained by replacement of the kaon mass by the pion mass: 
\begin{eqnarray}\label{eq:nunu}
\frac{d{\Gamma}}{dx}&=&\frac{1}{2^7{\pi}^5}G_F^2F^2f^2_{\pi}(1+r^2-2x)\sqrt{x^2-r^2} \nonumber \\
&&{\times}[(1-2x)x+r^2]
\end{eqnarray}
where $F$ is the hypothetical neutrino-neutrino interaction constant, $x=E_{\mu}/m_{\pi}$ ($x=E_{e}/m_{\pi}$), $r=m_{\mu}/m_{\pi}$ ($r=m_{e}/m_{\pi}$), $m_{\pi}$ and $m_{\mu}$ ($m_e$) are the masses of the pion and muon (positron), and $E_{\mu}$ ($E_e$) is the total muon (positron) energy. 

Another model with six-fermion interactions in addition to the usual four-fermion weak interactions was suggested by Ericson and Glashow \cite{6-fermion}. 
Vanzha, Isaev, and Lapidus \cite{6-fermion2} extended this to the four-body kaon decays. 
The equivalent differential decay rate \cite{Pang} was calculated for ${\pi}^+{\to}{\mu}^+{\nu_{\mu}}{\nu}\bar{\nu}$ as 
\begin{eqnarray}\label{eq:six}
\frac{d\Gamma}{dx}&=&\frac{m_\pi^9 f_\pi^2 F_S^2}{3\pi^2 2^5} (1-x)(x+r)(1+r^2-2x)^2 \nonumber \\
&&{\times}\sqrt{x^2-r^2}
\end{eqnarray}
where $F_S$ is the common form factor related to  $G_F$. 
For ${\pi}^+{\rightarrow}e^+{\nu_e}{\nu}\bar{\nu}$ decay, $m_\mu$ and $E_{\mu}$ were replaced by $m_e$ and $E_e$, respectively.

\begin{figure}
\includegraphics[scale=0.45]{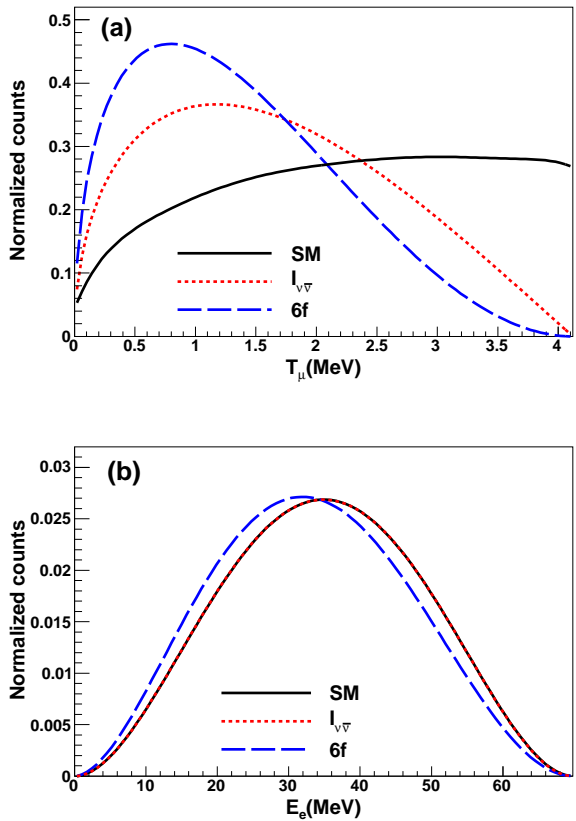}
\caption{\label{fig:EnergySpectra} The muon kinetic energy spectra of ${\pi}^+{\to}{\mu}^+{\nu_{\mu}}{\nu}\bar{\nu}$ decay (a) and the positron total energy for  ${\pi}^+{\to}e^+{\nu}_e{\nu}\bar{\nu}$ decay (b) for the SM (solid black), the neutrino-neutrino interaction ($I_{{\nu}\bar{\nu}}$, dotted red), and the six-fermion interaction (6f, dashed blue). The spectra in each panel are normalized to the same area for comparison.}
\end{figure}

The differential muon kinetic (positron total) energy spectrum of the rare pion decays ${\pi}^+{\to}{\mu}^+{\nu_{\mu}}{\nu}\bar{\nu}$ (${\pi}^+{\to}e^+{\nu_e}{\nu}\bar{\nu}$) for the neutrino-neutrino and six-fermion interaction models are presented in Fig. \ref{fig:EnergySpectra}. 
The SM and neutrino-neutrino interaction model spectra for ${\pi}^+{\to}e^+{\nu}_e{\nu}\bar{\nu}$ decay have similar shapes due to the small mass of the positron.

\section{Experiment}

\begin{figure}
\includegraphics[scale=0.4, clip, trim=7cm 0cm 7cm 0cm]{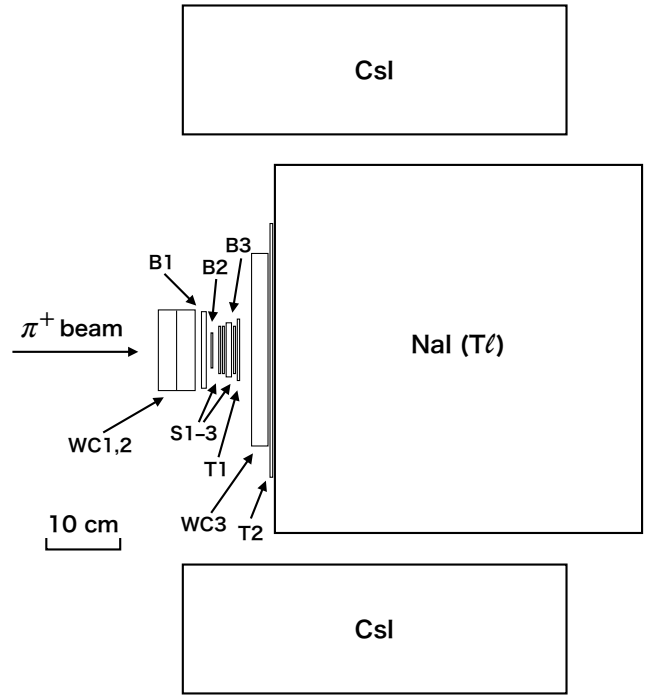}
\caption{\label{fig:detector} Schematic of the PIENU detector \cite{NIMA}.}
\end{figure}

The PIENU detector \cite{NIMA} illustrated in Fig. \ref{fig:detector} was designed to measure the pion branching ratio ${\Gamma}[{\pi}^+{\to}e^+{\nu_e}({\gamma})]/{\Gamma}[{\pi}^+{\rightarrow}{\mu}^+{\nu_{\mu}}({\gamma})]$ \cite{PIENU} where ($\gamma$) indicates the inclusion of radiative decays. 
The emitted positron in ${\pi}^+{\rightarrow}e^+{\nu_e}$ decay has total energy $E_e=69.8$ MeV. 
For ${\pi}^+{\to}{\mu}^+{\nu_{\mu}}$ decay followed by ${\mu}^+{\to}e^+{\nu_e}\bar{\nu_{\mu}}$ decay (${\pi}^+{\to}{\mu}^+{\rightarrow}e^+$ decay chain), the decay muon has kinetic energy $T_{\mu}=4.1$ MeV and a range in plastic scintillator of about 1 mm; the total energy of the positron in subsequent muon decay $\mu^+\rightarrow e^+ \nu_e\overline\nu_{\mu}$ ranges from $E_e=0.5$ to 52.8 MeV.

In the PIENU experiment, pions with momentum $75 \pm 1$ MeV/$c$ were provided by the TRIUMF M13 beam line \cite{M13} and tracked by two multiwire proportional chambers (WC1 and WC2) and two sets of silicon strip detectors (S1 and S2).
Two thin plastic scintillators (B1 and B2) were placed between WC2 and S1 to measure the time and energy loss for pion particle identification. 
Pions stopped and decayed at  rest in the center of an 8 mm thick plastic scintillator target (B3). 

Decay positrons emitted from B3 were tracked  by a silicon strip detector (S3) and a multiwire proportional chamber (WC3) placed downstream of the target. 
Two thin plastic scintillators (T1 and T2) were used to measure the time of decay positrons. 
The energies of decay positrons were measured by a 48 cm (dia.) $\times$ 48 cm (length) single crystal NaI(T$\ell$) calorimeter surrounded by 97 pure CsI crystals to detect shower leakage. 
The energy resolution of the calorimeter for positrons was 2.2\% (FWHM) at 70 MeV. 
A detailed description of the detector can be found in Ref. \cite{NIMA}. 

The pion and decay positron signals were defined by a coincidence of B1, B2, and B3, and a coincidence of T1 and T2, respectively. 
A coincidence of the pion and decay positron signals within a time window of $-$300 ns to 540 ns with respect to the pion signal was the basis of the main trigger condition. 
This was prescaled by a factor of 16 to form an unbiased trigger. 
 ${\pi}^+{\rightarrow}e^+{\nu_e}$ event collection was enhanced by an early time trigger selecting all events occurring from 6 to 46 ns after the arrival of the pion.  
The typical trigger rate including calibration triggers was about 600 s$^{-1}$. 

Plastic scintillators, silicon strip detectors and CsI crystals, and the NaI(T$\ell$) crystal were read out by 500 MHz, 60 MHz, and 30 MHz digitizers to extract the charge and time information of pulses. 
The wire chambers and trigger signals were read by multi-hit time$-$to$-$digital converters with 0.625 ns resolution.

\section{\label{muon}${\pi}^+{\rightarrow}{\mu}^+{\nu_{\mu}}{\nu}\bar{\nu}$ decay selection and analysis}

\begin{figure}
\includegraphics[scale=0.45]{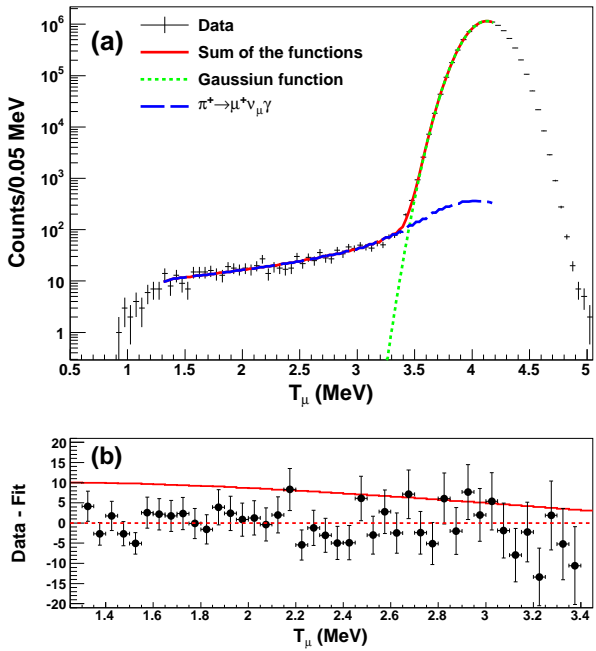}
\caption{\label{fig:FitH} (a) The $T_{\mu}$ spectra of ${\pi}^+{\to}{\mu}^+{\nu}_{\mu}$ decay. The black crosses with the statistical uncertainties show the data. The dotted green line, dashed blue line, and solid red line represent the Gaussian distribution at 4.1 MeV, ${\pi}^+{\rightarrow}{\mu}^+{\nu}_{\mu}{\gamma}$ decay, and the sum of those two functions, respectively. (b) Residual plot shown by the black circles with the statistical error bars for the signal region $T_{\mu}$=1.3 to 3.4 MeV. The solid red curve represents the hypothetical neutrino-neutrino interaction ($I_{{\nu}\bar{\nu}}$) signal with the branching ratio $B^{{\pi}{\mu}3{\nu}}_{I_{{\nu}\bar{\nu}}}=6.0{\times}10^{-5}$. The dashed horizontal red line indicates the residual of 0.}
\end{figure}

The signal of the rare pion decay ${\pi}^+{\rightarrow}{\mu}^+{\nu_{\mu}}{\nu}\bar{\nu}$ can be sought by examining the muon energy spectrum in ${\pi}^+{\rightarrow}{\mu}^+{\rightarrow}e^+$ decay. 
The cuts used for the analysis were the same as for the heavy neutrino search \cite{PIENU2}. 
Pions were identified using the energy loss information in B1 and B2 and events with extra hits in B1, B2, T1 and T2 were rejected. 

To ensure that the events selected were from ${\pi}^+{\rightarrow}{\mu}^+{\rightarrow}e^+$ decays, late positron decay time $t>200$ ns after the pion stop, a solid angle fraction of about 20\% determined by the position of hits in WC3 for the decay positron track, and the positron energy in the NaI(T$\ell$) calorimeter $E_e<55$ MeV were required. 
Then, the events with three clearly separated pulses in the target (B3) were selected and the second pulse information was extracted and assigned to the decay muon \cite{PIENU2}.
The muon kinetic energy ($T_{\mu}$) spectrum after the event selection cuts is shown in Fig. \ref{fig:FitH} (a). 
The drop below 1.2 MeV was due to the inefficiency of the pulse detection logic \cite{PIENU2}. 
The main background below 3.4 MeV was due to the radiative pion decay ${\pi}^+{\rightarrow}{\mu}^+{\nu_{\mu}}{\gamma}$ (branching fraction $2{\times}10^{-4}$ \cite{pimunug}). 
The total number of ${\pi}^+{\to}{\mu}^+{\to}e^+$ events available was 9.1${\times}10^6$. 

The decay ${\pi}^+{\rightarrow}{\mu}^+{\nu}_{\mu}{\nu}\bar{\nu}$ was searched for by fitting the $T_{\mu}$ energy spectrum of ${\pi}^+{\to}{\mu}^+{\to}e^+$ decays. 
The fit was performed using a Gaussian peak centered at 4.1 MeV (energy resolution ${\sigma}=0.16$ MeV), the ${\pi}^+{\rightarrow}{\mu}^+{\nu}_{\mu}{\gamma}$ decay spectrum obtained by Monte Carlo (MC) simulation \cite{geant4}, and the normalized signal spectra shown in Fig. \ref{fig:EnergySpectra} (a) including the energy resolution in B3. 
The fit for $T_{\mu}$ from 1.3 to 4.2 MeV without any ${\pi}^+{\rightarrow}{\mu}^+{\nu_{\mu}}{\nu}\bar{\nu}$ signal introduced gave  ${\chi}^2/$dof=1.27 (dof=53) and the residuals of the fit for the signal sensitive region are shown in Fig. \ref{fig:FitH} (b). 
The addition of signal components did not change the fit result. 

No significant signal beyond the statistical uncertainty was observed. 
For example, the SM signal branching ratio obtained by the fit was $B^{{\pi}{\mu}3{\nu}}_{\rm SM}=(-9.4{\pm}9.7){\times}10^{-6}$. 
Systematic uncertainties and acceptance effects were approximately canceled by taking the ratio of amplitudes for the signal and ${\pi}^+{\to}{\mu}^+{\nu}_{\mu}$ decays. 
The following upper limits for ${\pi}^+{\rightarrow}{\mu}^+{\nu_{\mu}}{\nu}\bar{\nu}$ decay with the SM, neutrino-neutrino, and six-fermion interactions calculated with the Feldman-Cousins (FC) approach \cite{FC} were found:
\begin{eqnarray}
B^{\pi\mu3\nu}_{\rm SM}&<&8.6{\times}10^{-6}, \\
B^{\pi\mu3\nu}_{I_{{\nu}\bar{\nu}}}&<&6.4\times10^{-6},~{\rm and}\\
B^{\pi\mu3\nu}_{\rm 6f}&<&6.2{\times}10^{-6}.
\end{eqnarray}
These are the first results reported for ${\pi}^+{\rightarrow}{\mu}^+{\nu_{\mu}}{\nu}\bar{\nu}$ decay.

\section{${\pi}^+{\rightarrow}e^+{\nu_e}{\nu}\bar{\nu}$ decay selection and analysis}

\begin{figure*}
\includegraphics[scale=0.9]{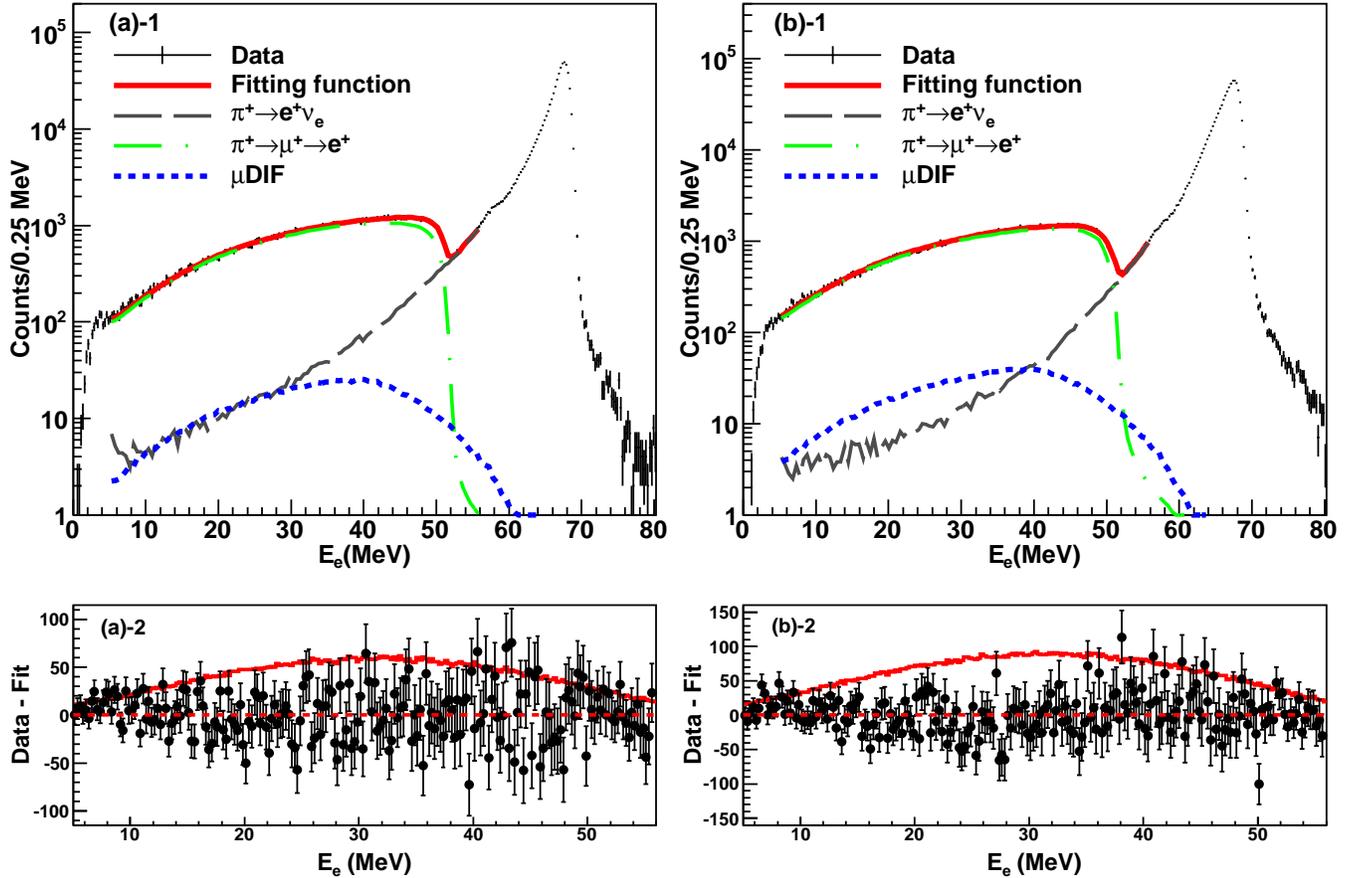}
\caption{\label{fig:Suppressed} Top: The $E_e$ spectra of ${\pi}^+{\to}e^+{\nu_e}$ decay after ${\pi}^+{\rightarrow}{\mu}^+{\rightarrow}e^+$ suppression cuts from data taken before (a)-1 and after (b)-1 November 2010. The black crosses with the statistical uncertainties show the data. Background components illustrated by the dashed and dotted green line, dashed gray line, dotted blue line, and solid red line represent ${\pi}^+{\rightarrow}{\mu}^+{\rightarrow}e^+$ decays, low energy ${\pi}^+{\rightarrow}e^+{\nu_e}$ tail, $\mu$DIF events, and the sum of those three components, respectively (see text). Bottom: The residual plots shown by the black circles with the statistical error bars and hypothetical neutrino-neutrino interaction ($I_{ {\nu}\bar{\nu} }$) signals (solid red curves) with the branching ratio $B^{{\pi}e3{\nu}}_{I_{{\nu}\bar{\nu}}}=2.0{\times}10^{-6}$ for data taken before (a)-2 and after (b)-2 November 2010. The dashed horizontal red lines represent the residual of 0.}
\end{figure*}

Because the calibration system for the  CsI crystals was not available before November 2010, the data for the ${\pi}^+{\rightarrow}e^+{\nu_e}{\nu}\bar{\nu}$ decay analysis were divided into two sets. 
A 15\%  solid angle cut was applied to the data taken after November 2010, and a tighter cut (10\%) was used for the data taken before November 2010 to minimize the effects of electromagnetic shower leakage.
The cuts used for the pion selection and the rejection of the extra activity are the same as described in Sec. \ref{muon}. 

For the ${\pi}^+{\rightarrow}e^+{\nu}_e{\nu}\bar{\nu}$ decay study, the ${\pi}^+{\rightarrow}{\mu}^+{\rightarrow}e^+$ backgrounds were suppressed using decay time, energy, and tracking information provided by WC1, WC2, S1, and S2 \cite{PIENU3}. 
The ${\pi}^+{\rightarrow}{\mu}^+{\rightarrow}e^+$ suppression cuts were based on the heavy neutrino analysis \cite{PIENU3} but optimized for this analysis to minimize distortion in the ${\pi}^+{\rightarrow}e^+{\nu}_e$ energy spectrum.
The decay times were required to be from $t=$ 7 to 35 ns after the pion stop to exploit the short pion lifetime compared to the muon lifetime. 
For ${\pi}^+{\rightarrow}{\mu}^+{\rightarrow}e^+$ decay, the energy deposit in B3 was 4.1 MeV larger than for ${\pi}^+{\rightarrow}e^+{\nu_e}$ decay due to the presence of the muon. 
After the target energy cut, a beam pion tracking cut which provided the incoming angle was applied to reject pion decay-in-flight events before the target \cite{NIMA}. 
Figure \ref{fig:Suppressed} shows the decay positron energy spectra of ${\pi}^+{\rightarrow}e^+{\nu_e}$ decays after ${\pi}^+{\rightarrow}{\mu}^+{\rightarrow}e^+$ background suppression cuts. 
The bump in the decay positron energy spectra at around 58 MeV was due to photo-nuclear reactions in the NaI(T$\ell$) \cite{PN}. 
The total number of ${\pi}^+{\rightarrow}e^+{\nu_e}$ events was $1.3{\times}10^6$. 

The decay ${\pi}^+{\rightarrow}e^+{\nu_e}{\nu}\bar{\nu}$ was searched for by fitting the background-suppressed decay positron energy spectrum. 
The background component due to the remaining ${\pi}^+{\rightarrow}{\mu}^+{\rightarrow}e^+$ events was obtained from the data by requiring a late time $t>200$ ns. 
The shape of the low energy ${\pi}^+{\rightarrow}e^+{\nu_e}$ tail was obtained by MC simulation including the detector response and radiative decay. 
Another background came from the decays-in-flight of muons ($\mu$DIF) following ${\pi}^+{\rightarrow}{\mu}^+{\nu_{\mu}}$ decays in B3 that has similar time and energy distributions to  ${\pi}^+{\rightarrow}e^+{\nu_e}$ decay. 
The shape of the $\mu$DIF event spectrum was obtained by MC simulation. 
The signal shapes shown in Fig. \ref{fig:EnergySpectra} (b) including the detector response were normalized to 1 and used for the fit. 
To combine the two data sets, a common branching ratio was used as a free parameter in the fit. 
The fit in the range of $E_e=5$ to $56$ MeV without any signal resulted in  ${\chi}^2/$dof=1.04 (dof=402). 
Addition of ${\pi}^+{\rightarrow}e^+{\nu_e}{\nu}\bar{\nu}$ signal shapes did not change the fit result. 

No significant excess above the statistical uncertainty was observed. 
For example, the branching ratio assuming the SM obtained by the fit was $B^{{\pi}e3{\nu}}_{\rm SM}=(-1.8{\pm}1.9){\times}10^{-7}$. 
The statistical uncertainty is dominant because the systematic uncertainties and the acceptance effects are approximately canceled out by taking the ratio of the number of the signal events obtained by the fit to the number of pion decays. 
The upper limits for the branching ratio ${\pi}^+{\rightarrow}e^+{\nu_e}{\nu}\bar{\nu}$ were determined using the FC approach: 
\begin{eqnarray}
B^{\pi e 3\nu}_{\rm SM}&<&1.6{\times}10^{-7}, \\
B^{\pi e 3\nu}_{I_{{\nu}\bar{\nu}}}&<&1.6{\times}10^{-7},~{\rm and} \\
B^{\pi e 3\nu}_{\rm 6f}&<&1.5{\times}10^{-7}.
\end{eqnarray}
Compared to the previous TRIUMF experiment \cite{Picciotto}, the limits were improved by an order of magnitude.

\section{Summary}

No evidence of the rare pion decays ${\pi}^+{\to}{\mu}^+{\nu_{\mu}}{\nu}\bar{\nu}$ and ${\pi}^+{\to}e^+{\nu_e}{\nu}\bar{\nu}$ was found and new upper limits were set using the SM and non-SM neutrino-neutrino and six-fermion interactions. 
For ${\pi}^+{\rightarrow}{\mu}^+{\nu_{\mu}}{\nu}\bar{\nu}$ decay, the limits obtained are the first available results. 
The limits on the branching ratio for ${\pi}^+{\rightarrow}e^+{\nu}_e{\nu}\bar{\nu}$ decay were improved by an order of magnitude.

\begin{acknowledgments}
This work was supported by the Natural Sciences and Engineering Research Council of Canada (NSERC, No. SAPPJ-2017-00033), and by the Research Fund for the Doctoral Program of Higher Education of China, by CONACYT doctoral fellowship from Mexico, and by JSPS KAKENHI Grant No. 18540274, No. 21340059, No. 24224006, and No. 19K03888 in Japan.
We are grateful to Brookhaven National Laboratory for the loan of the crystals, and to the TRIUMF operations, detector, electronics and DAQ groups for their engineering and technical support.
\end{acknowledgments}

\appendix
\section{The SM calculation}

\subsection{Squared amplitudes}

The SM decay rates for  ${\pi}^+{\to}l^+{\nu}_l{\nu}\bar{\nu}, l=e, \mu$ were obtained  based on the calculation of the rare kaon decay ${K}^+{\to}l^+{\nu}_l{\nu}\bar{\nu}$ \cite{3nu}.
The sum of the amplitudes for  ${\pi}^+{\to}{\mu}^+{\nu}_{\mu}{\nu}\bar{\nu}$ decay for all six diagrams  in Fig. \ref{fig:Feynman}  can be written as 
\begin{equation}
\label{amplitude}
{\cal M}=\frac{f_{\pi}G_F^2V_{ud}}{\sqrt{2}} \times \sum_{i=1}^6 M_i \equiv
\frac{f_{\pi}G_F^2V_{ud}}{\sqrt{2}} \times M
\end{equation}
where $M_i$ is the amplitude of each process represented as
\begin{widetext}
\begin{eqnarray}
M_1&=&2 \sin^2 \theta_W \cdot \bar \nu_l (p_1) \gamma^{\lambda}
 (1 - \gamma_5) \nu_{l} (p_2) 
 \cdot \bar \nu_{\mu}(p_3) \gamma_{\lambda} (
 1-\gamma_5 ) \mu(p_4) \\
M_2&=&(1-2 \sin^2
 \theta_W) \cdot \dfrac{(p_3+p_4)^{\lambda} (p_0 + p_3+p_4)^{\rho}}{(p_3+p_4)^2 -
   m_\pi^2} 
   \cdot \bar \nu_l (p_1) \gamma_{\rho} (1 - \gamma_5) \nu_{l}
 (p_2) \cdot \bar \nu_{\mu}(p_3) \gamma_{\lambda} ( 1-\gamma_5 ) \mu(p_4)
 \\
M_3&=& - \frac{1}{\sqrt{2}} \cdot \dfrac{(p_1+p_2)^{\lambda} (p_0
   +p_1+p_2)^{\rho}}{(p_1+p_2)^2 - m^2_{\pi}} 
   \cdot \bar \nu_l (p_1)
 \gamma_{\lambda} (1 - \gamma_5) \nu_{l} (p_2) \cdot \bar \nu_{\mu}(p_3)
 \gamma_{\rho} ( 1-\gamma_5 ) \mu(p_4) \\
M_4&=& \frac{1}{2} \cdot
 \dfrac{1}{(p_0-p_4)^2} p^\rho_{0} \cdot  \bar \nu_l (p_1)
 \gamma^{\lambda} (1 - \gamma_5) \nu_{l} (p_2) 
 \cdot \bar \nu_{\mu}(p_3)
 \gamma_{\lambda} ( 1-\gamma_5 ) (\hat p_0 - \hat p_4) \gamma_{\rho}
 (1-\gamma_5) \mu(p_4) \\
M_5&=& \frac{1}{2} \cdot p^\lambda_{0}
  \bar \nu_{\mu}(p_3) \gamma_{\lambda} (1 - \gamma_5) \dfrac{ - \hat p_0 +
   \hat p_3 + m_{\mu}}{(p_0-p_3)^2 - m_{\mu}^2} \gamma^{\rho} 
   (4
 \sin^2 \theta_W - 1 + \gamma_5) \mu(p_4) \cdot \bar \nu_l (p_1)
 \gamma_{\rho} (1 - \gamma_5) \nu_{l} (p_2) \\
M_6&=& p^\lambda_{0}  \cdot \bar \nu_{l} \gamma_{\lambda} (1-\gamma_5) \dfrac{m_l - \hat
   p_0 + \hat p_1}{(p_0-p_1)^2 - m_l^2} \gamma^{\rho}(1-\gamma_5)
 \nu_{l}(p_2) 
 \cdot \bar \nu_{\mu}(p_3) \gamma_{\rho} (1-\gamma_5)
 \mu(p_4).
\end{eqnarray}
\end{widetext}
In the equation above, $p_i$ is the momentum of the $i^{th}$ outgoing particle ($i=1,2,3,4$), $m_{\pi}$ and $m_{\mu}$ are the pion and muon masses,  $l=e,\mu,\tau$, $\hat{p_i}=p_{i_\mu} \gamma^\mu$, and $\gamma^\mu$  are the Dirac matrices. 
Similar amplitudes contribute to the decay ${\pi}^+{\to}e^+{\nu}_e{\nu}\bar{\nu}$ with replacements of ${\mu}(p_4)$ to $e(p_4)$, ${\nu}_{\mu}$ to ${\nu}_e$, and $m_{\mu}$ to $m_e$. 
For the decay ${\pi}^+{\to}e^+{\nu}_e{\nu}_{\mu}\bar{\nu}_{\mu}$, amplitude $M_6$  has a resonance divergence associated with an on-shell muon that does not apply when the positron is produced directly. 
To calculate the non-resonant contribution, this amplitude was excluded from the ${\pi}^+{\to}e^+{\nu}_e{\nu}_{\mu}\bar{\nu}_{\mu}$ decay calculation. 

The squared  matrix element describing the four body decay ${\pi}^+{\to}{\mu}^+{\nu}_{\mu}{\nu}\bar{\nu}$ is presented using the notation for the scalar product of four-vectors $p_i$ and $p_j$,  $p_ip_j{\equiv}x_{ij}$ ($i<j$).  Then, the corresponding squared matrix element for ${\pi}^+{\to}{\mu}^+{\nu}_{\mu}{\nu}\bar{\nu}$ decay into the final states with electron and tau neutrinos is 
\begin{widetext}
\begin{eqnarray}
{M}^2 &=& \big|A \times \bar \nu_l (p_1) \gamma^{\mu} (1 - \gamma_5)
\nu_{l} (p_2) \cdot\bar \nu_{\mu}(p_3) \gamma_{\mu} ( 1-\gamma_5 )
\mu(p_4) B \times \bar \nu_l (p_1) \hat p_0 (1 - \gamma_5) \nu_l (p_2)  \times \bar \nu_{\mu} (p_3) (1 + \gamma_5)\mu(p_4) \nonumber \\
&&+ C \times \bar \nu_l (p_1) \gamma_{\mu}
(1-\gamma_5) \nu (p_2)\cdot\bar \nu_{\mu} (p_3) \gamma_{\mu}(1-\gamma_5) \hat p_0 \mu(p_4)\big|^2 \nonumber \\
&=& 256 A^2 x_{13}x_{24} + 64 B^2 x_{34} \l 2x_{01}x_{02} - x_{12} m_\pi^2 \r + 256 C^2
\l 2 x_{13}x_{02} x_{04} - m_\pi^2 x_{13} x_{24}\r \nonumber \\
&&- 128 A B m_{\mu} \l x_{13}x_{02} +x_{01}x_{23} -x_{12}x_{03} \r - 512m_{\mu} A C x_{13}x_{02} \nonumber \\
&&+ 128 B C \left[ 2 x_{02} \l x_{01}x_{34} \right.\right. + \left.\left. x_{13}x_{04} - x_{03}x_{14} \r- m_\pi^2 \l x_{12}x_{34} +x_{13}x_{24} -x_{14}x_{23}\r \right]\,
\end{eqnarray}
where
\begin{eqnarray}
A &=& 2 \l\sin^2 \theta_W+ \frac{m_\pi^2 - 2x_{04}}{2  (m_\pi^2 - 2x_{04} + m^2_{\mu})}- \frac{m_\pi^2 - 2x_{01}}{ m_\pi^2 - 2x_{01} - m_{l}^2}\right. \nonumber \\
&&+\left.\l \frac{1-2\sin^2\theta_W}{2}\r \l 1+\frac{m_\mu^2}{2x_{12}+2x_{14}+2x_{24}}\r \r, \nonumber \\
B &=& -2 m_{\mu} \l \dfrac{1 - 2 \sin^2 \theta_W}{2x_{12} - 2x_{01} - 2x_{02}} -\dfrac{2\sin^2\theta_W}{2x_{12}+2x_{14}+2x_{24}}\r,~{\rm and} \nonumber \\
C &=& - m_{\mu} \l \dfrac{1}{m_\pi^2 -2x_{04} + m_{\mu}^2} + \dfrac{2\sin^2\theta_W}{2x_{12}+2x_{14}+2x_{24}} \r. \nonumber
\end{eqnarray}
The squared matrix element for ${\pi}^+{\to}{\mu}^+{\nu}_{\mu}{\nu}\bar{\nu}$ decays with muon neutrinos in
the final state is
\begin{eqnarray}
{M}^2 &=& \big|A \times \bar \nu_l (p_1) \gamma^{\lambda} (1 - \gamma_5)
\nu_{l} (p_2) \cdot \bar \nu_{\mu}(p_3) \gamma_{\lambda} ( 1-\gamma_5 )
\mu(p_4) + B \times \bar \nu_l (p_1) \hat p_0 (1 - \gamma_5) \nu_l
(p_2) \nonumber \\ 
&& \times \bar \nu_{\mu} (p_3) (1 + \gamma_5)\mu(p_4) + C \times \bar \nu_l (p_1) \gamma_{\lambda}
(1-\gamma_5) \nu (p_2) \cdot \bar \nu_{\mu} (p_3) \gamma^{\lambda}(1-\gamma_5)
\hat p_0 \mu(p_4) \nonumber \\
&& + D \times \bar \nu_l (p_3) \hat p_0 (1 -
\gamma_5) \nu_l (p_2)\cdot\bar \nu_{\mu} (p_1) (1 + \gamma_5)\mu(p_4) +
E \times \bar \nu_l (p_3) \gamma_{\lambda} (1-\gamma_5) \nu (p_2) \nonumber \\ 
&& \times \bar \nu_{\mu} (p_1) \gamma^{\lambda}(1-\gamma_5) \hat p_0
\mu(p_4)\big|^2 \nonumber \\
&=& 256 A^2 x_{13}x_{24} + 64 B^2 x_{34}
\l 2x_{01}x_{02} - x_{12} m_\pi^2\r  + 256 C^2 x_{13} \l 2 x_{02}x_{04} -
m_\pi^2 x_{24}\r \nonumber \\
&&- 128 A B m_{\mu} \l x_{13}x_{02} + x_{01}x_{23} -x_{12}x_{03} \r - 512m_{\mu} A C x_{13}x_{02} + 128 B
C \left[ 2 x_{02} \l x_{01}x_{34}  \right. \right. \nonumber \\
&&+ \left.\left. x_{13}x_{04} - x_{03}x_{14} \r-
m_\pi^2 \l x_{12}x_{34} +x_{13}x_{24} -x_{14}x_{23} \r \right] + 64 D^2 x_{14} \l
2 x_{03} x_{02} - m_\pi^2 x_{23} \r \nonumber \\
&&+ 256 E^2 x_{13} \l 2 x_{02} x_{04}
- m_\pi^2 x_{24} \r - 128 m_{\mu} A D \l x_{12} x_{03} + x_{13}
x_{02} - x_{01} x_{23}\r -512 m_{\mu} A E \nonumber \\
&& \times  x_{13} x_{02} - 64 B D \left[ 2x_{02} \l x_{01}x_{34} + x_{14}x_{03} - x_{13}x_{04}\r  -
m_\pi^2 \l x_{12}x_{34} + x_{14}x_{23} - x_{13} x_{24}\r \right] \nonumber \\
&&+128 B E \left[2 x_{02} \l x_{01} x_{34} + x_{13}x_{04} - x_{14}x_{03}\r - m_\pi^2
\l x_{12}x_{34} + x_{13}x_{24} - x_{14}x_{23}\r \right] \nonumber \\
&&+128 C D \left[ 2 x_{02}\l x_{03} x_{14} + x_{13}x_{04} - x_{34}x_{01}\r - m_\pi^2
\l x_{23}x_{14} + x_{13}x_{24} - x_{34}x_{12}\r \right] \nonumber \\
&&+512 C E x_{13} \l 2x_{04}x_{02} - m_\pi^2 x_{24}\r  + 128 E D \left[2 x_{02} \l x_{03}x_{14} +
x_{13}x_{04} - x_{34}x_{01}\r \right. \nonumber \\
&&- \left. m_\pi^2 \l x_{23}x_{14} +  x_{13}x_{24} - x_{34}x_{12}\r \right]
\end{eqnarray}
with
\begin{eqnarray}
A &=& 2  \left[ \l 2\sin^2 \theta_W + \frac{m_\pi^2 -2 x_{04}}{
	m_\pi^2 - 2x_{04} +m^2_{\mu}} \r + \frac{ 1 - 2 \sin^2 \theta_W}{2}
\l 2+ \frac{m_{\mu}^2}{2x_{12} + 2x_{14} + 2x_{24}}\right.\right. \nonumber \\
&&+ \left. \left. \frac{m_{\mu}^2}{2x_{23} + 2x_{34} +2x_{24}} \r -
\l \frac{m_\pi^2 - 2x_{01}}{m_\pi^2 - 2x_{01} -
	m_{\mu}^2} + \frac{m_\pi^2 - 2x_{03}}{m_\pi^2 - 2x_{03}- m_{\mu}^2} \r \right] , \nonumber \\
B &=& 2  m_{\mu} \l \frac{2 \sin^2 \theta_W -
	1}{2x_{12} - 2x_{01} - 2x_{02}} + \frac{2 \sin^2 \theta_W}{2x_{12}
	+2x_{14} +2x_{24}} \r , \nonumber \\
C &=& - 2 m_{\mu}\l\frac{1}{2(m_\pi^2 - 2x_{04} +m_{\mu}^2)} + \frac{\sin^2
	\theta_W}{2x_{12} + 2 x_{14} + 2 x_{24}} \r , \nonumber \\
D &=& - 2 m_{\mu} \l \frac{2 \sin^2 \theta_W - 1}{2x_{23} - 2x_{03} -
	2x_{02}} + \frac{2 \sin^2 \theta_W}{2x_{23} +2x_{34} +2x_{24}} \r ,~{\rm and} \nonumber \\
E &=& 2 m_{\mu} \l \frac{1}{2(m_\pi^2 - 2x_{04} +m_{\mu}^2)} + \frac{\sin^2 \theta_W}{2x_{23} + 2 x_{34} + 2x_{24}} \r . \nonumber
\end{eqnarray}

\subsection{Differential decay rate and branching fraction}

The differential decay rate for ${\pi}^+{\to}l^+{\nu}_l{\nu}\bar{\nu}$ can be represented using Eqs. (B.1) to (B.3) in Ref. \cite{3nu} as
\begin{eqnarray}
\frac{d {\bf \Gamma}^{{\pi}l3{\nu}}}{d{\bf p}_4}  &=& \frac{1}{(2 \pi)^6} \int^{M_3^2}_{0} d M_2^2 \int_{-1}^{1} d \cos \theta_1 \int_{-1}^{1} d \cos \theta_2 \int_{0}^{2 \pi} d \phi \nonumber \\ 
&&\times \frac{[(m_\pi^2 - (M_3 + m_{4})^2)(m_\pi^2 -(M_3 - m_{4})^2)]^{1/2} \times \mathbf{p}_3^2 \times \mathbf{p}_2^2}{2m_\pi^2\times (E_{123}\mathbf{p}_3 - \mathbf{p}_{123}E_3 \cos\theta_1)(E_{12}\mathbf{p}_2 - \mathbf{p}_{12}E_2 \cos\theta_2)}\, \times \frac{\textbf{p}_4  \left| {\cal M} \right|^2}{32 E_4} \,.
\end{eqnarray}
Here, we adopt the  auxiliary momentum variables $p_{12}=(E_{12},\vec{p}_{12})=p_1+p_2$ and $p_{123}=(E_{123},\vec{p}_{123})=p_1+p_2+p_3$, so that $M_{2}^2=p_{12}^2$, $M_3^2=p_{123}^2$,  angle variables ${\theta}_1{\equiv}{\angle}(\vec{p}_3,\vec{p}_{123})$ and ${\theta}_2 {\equiv} {\angle}(\vec{ p}_2,\vec p_{12}) $, and $\phi$ is the rotation angle of the plane $(\vec{p}_{12}, \vec{p}_2)$ around $\vec{p}_{12}$.
The branching fraction is defined as 
\begin{equation}
B^{{\pi}l3{\nu}}_{SM}{\equiv}{\tau}_{\pi}{\Gamma}^{{\pi}l3{\nu}}
\end{equation}
where ${\tau}_{\pi}$ is the pion lifetime.

\subsection{Results}

For the results presented below, the neutrinos are treated as massless particles; 
then, the amplitude $M_3$ which refers to diagram 3 in Fig. 1  vanishes. 
The differential momentum and energy spectra of muons in
${\pi}^+{\to}{\mu}^+{\nu}_{\mu}{\nu}\bar{\nu}$ decay and positrons in ${\pi}^+{\to}e^+{\nu}_{e}{\nu}\bar{\nu}$ decay are illustrated in Fig. \ref{fig:momentum} and \ref{fig:EnergySpectra}.

\begin{figure}
%\begin{center}
\includegraphics[width=17cm]{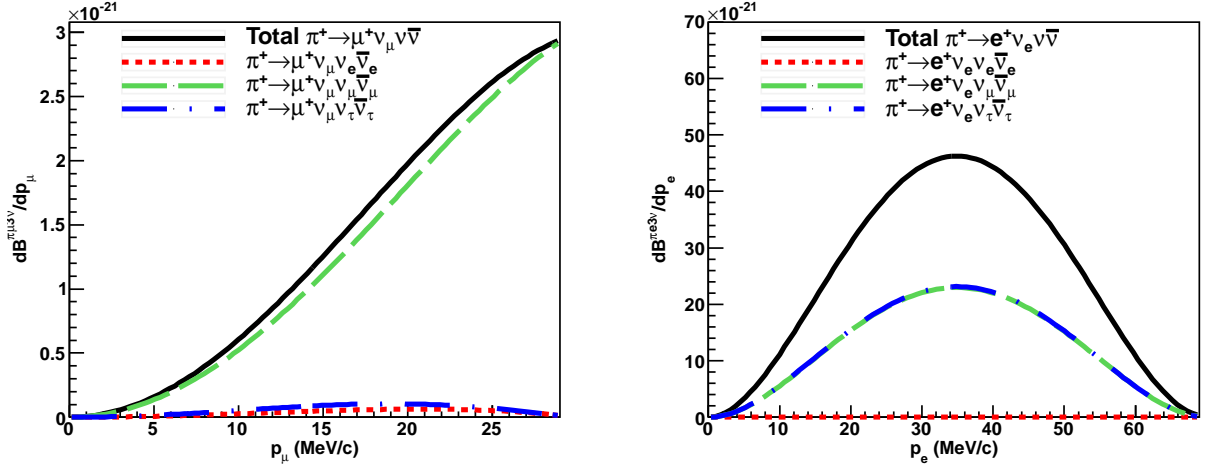}
\caption{\label{fig:momentum} The SM differential momentum spectra of  muons (left) and positrons (right) in ${\pi}^+{\to}l^+{\nu}_l{\nu}\bar{\nu}$ decays.}
%\end{center}
\end{figure}

For the decay ${\pi}^+{\to}{\mu}^+{\nu}_{\mu}{\nu}\bar{\nu}$, the polynomials were built on the interval $ \mathbf{p}_\mu = [0; 29.6] ~ {\rm MeV}/c, \ ({\mathbf{p}_\mu}_{\rm max} \approx 29.8$ MeV/$c$). 
By making use of the interpolating polynomials in the lepton 3-momenta $p_{\mu}$($p_{e}$) [MeV/$c$]  numerical fits to these distributions were found:

\begin{equation}\label{dBmup}
\begin{split}
\frac{d \mathbf{B}^{\pi_{\mu 3\nu}}}{d \bf{p_{\mu}}} =  &-1.68\times 10^{-25}+3.50 \times 10^{-25} \mathbf{p_{\mu}} + 6.32 \times 10^{-24} \mathbf{p_{\mu}}^2 + 2.49 \times 10^{-26} \mathbf{p_{\mu}}^3
\\ &- 6.23 \times 10^{-27} \mathbf{p_{\mu}}^4 + 8.72 \times 10^{-29} \mathbf{p_{\mu}}^5 - 6.26 \times 10^{-31} \mathbf{p_{\mu}}^6\, ,
\end{split}
 \end{equation}
\begin{equation}\label{dBep}
\begin{split}
\frac{d \mathbf{B}^{\pi_{e 3\nu}}}{d \mathbf{p}_{e}} =  &-9.37 \times 10^{-25} + 1.14 \times 10^{-23} \mathbf{p}_{e} + 1.50 \times 10^{-22} \mathbf{p}_{e}^2 - 4.21 \times 10^{-24} \mathbf{p}_{e}^3 \\ 
& +2.71 \times 10^{-26} \mathbf{p}_{e}^4 + 5.46 \times 10^{-29} \mathbf{p}_{e}^5 - 2.68 \times 10^{-31} \mathbf{p}_{e}^6.
\end{split}
\end{equation}
The differential muon kinetic ($T_{\mu}$) and positron total ($E_e$) energy distributions are shown in Fig. \ref{fig:EnergySpectra}  and the interpolating polynomials are given below:
 \begin{equation}\label{eq:Bmu}
\begin{split}
 \frac{d \mathbf{B}^{\pi_{\mu 3\nu}}}{d T_\mu} =& 1.93\times 10^{-21}+1.48 \times 10^{-20} T_{\mu} - 1.62 \times 10^{-20} T_\mu^2 + 1.39 \times 10^{-20} T_{\mu}^3 - 8.09 \times 10^{-21} T_{\mu}^4 \\ 
& + 3.00 \times 10^{-21} T_{\mu}^5 - 6.79 \times 10^{-22} T_{\mu}^6 + 8.53 \times 10^{-23} T_{\mu}^7 - 4.55 \times 10^{-24} T_{\mu}^8,~{\rm and}
\end{split}
\end{equation}
\begin{equation}\label{eq:Be}
\begin{split}
\frac{d \mathbf{B}^{\pi_{e 3\nu}}}{d E_e} =& -6.80 \times 10^{-23} + 3.40 \times 10^{-23} E_e + 1.47 \times 10^{-22} E_e^2 - 4.04 \times 10^{-24} E_e^3 \\
& + 2.27 \times 10^{-26} E_e^4 + 1.08 \times 10^{-28} E_e^5 - 5.21 \times 10^{-31} E_e^6.
\end{split}
\end{equation}

By integration over the individual muon momentum spectrum for each $\nu \bar{\nu}$ pair, the following branching ratios were obtained:
\begin{displaymath}
\begin{array}{rclcrcl}
{\rm \mathbf{B}}(\pi^+ \rightarrow \mu^+ \nu_{\mu} \nu_{\mu} \bar \nu_{\mu} )
  &= & 3.7 \times 10^{-20}, & \hspace{1.5cm}

\\ {\rm \mathbf{B}}(\pi^+ \rightarrow \mu^+ \nu_{\mu} \nu_{e} \bar \nu_{e} ) & = &
1.0\times 10^{-21},~{\rm and} && 
\\ {\rm \mathbf{B}}(\pi^+ \rightarrow \mu^+ \nu_{\mu} \nu_{\tau} \bar \nu_{\tau}
) &=&1.7\times 10^{-21}. && 
\\ 
\end{array}
\end{displaymath}
Then the result for the summed branching ratio is $\mathbf{B}^{\pi \mu 3 \nu} = 4.0 \times 10^{-20}$. 

Similarly, integration over the individual positron momentum spectrum for each $\nu \bar{\nu}$ pair results in  the following branching ratios: 
\begin{displaymath}
\begin{array}{rclcrcl}
{\rm \mathbf{B}}(\pi^+ \rightarrow e^+ \nu_{e} \nu_{\mu} \bar \nu_{\mu} )
& = &8.6 \times 10^{-19},
\\ {\rm \mathbf{B}}(\pi^+ \rightarrow e^+ \nu_{e} \nu_{e} \bar
\nu_{e} ) & = & 6.1 \times 10^{-24},~{\rm and}
\\{\rm \mathbf{B}}(\pi^+ \rightarrow e^+ \nu_{e} \nu_{\tau} \bar \nu_{\tau} ) &=& 8.6 \times 10^{-19}.
\end{array}
\end{displaymath}
The summed branching ratio is $ \mathbf{B}^{\pi e 3 \nu} = 1.7 \times 10^{-18}$. 
The uncertainties on the branching ratios $ \mathbf{B}^{\pi \mu 3 \nu}$ and $ \mathbf{B}^{\pi e 3 \nu}$ were estimated to be $<1$\%.

\end{widetext}

\end{document}